\documentclass{PoS}

\newcommand{\beq}{\begin{equation}}   
\newcommand{\eeq}{\end{equation}}   
\newcommand{\beqn}{\begin{eqnarray}}   
\newcommand{\eeqn}{\end{eqnarray}}   
\newcommand{\nn}{\nonumber}

\title{O($a^2$) cutoff effects in Wilson fermion simulations}

\ShortTitle{O($a^2$) cutoff effects in Wilson fermion simulations}

\author{Roberto Frezzotti and \speaker{Giancarlo Rossi}%
         \thanks{On the behalf of the ETM Collaboration.}\\
        Univ. and INFN of Rome Tor Vergata, Via della\ Ricerca Scientifica 1, I-00133, Rome, Italy\\
        E-mail: \email{rossig@roma2.infn.it}\\
   \qquad\quad\,\,\email{frezzotti@roma2.infn.it}}

\abstract{We show that the size of the O($a^2$) flavour violating cutoff artifacts 
that have been found to affect the value of the neutral pion mass in 
simulations with maximally twisted Wilson fermions is controlled by a continuum 
QCD quantity that is fairly large and is determined by the dynamical 
mechanism of spontaneous chiral symmetry breaking. One can argue 
that the neutral pion mass is the only physical quantity blurred by such 
cutoff effects. O($a^2$) corrections of this kind are also present in  
standard Wilson fermion simulations, but they can either affect the determination 
of the pion mass or be shifted from the latter to other observables, 
depending on the way the critical mass is evaluated.}

\FullConference{The XXV International Symposium on Lattice Field Theory\\
                 July 30 - August 4 2007\\
                 Regensburg, Germany}

\begin{document}
\section{Introduction and main results}   
\label{sec:INTRO}   

Numerical data for the mass of the neutral pion in maximally twisted Lattice QCD 
(Mtm-LQCD)~\cite{TM} simulations show unnaturally large lattice 
artifacts~\cite{LET}~\footnote{The neutral to charged pion mass splitting 
measured in the unquenched Mtm-LQCD simulations carried out in ref.~\cite{LET} 
with the tree-level improved Symanzik gauge action turns out to be smaller 
(and of opposite sign) than the quenched result~\cite{XLF} where the standard 
plaquette gauge action was used. This finding is interesting in view of the 
established relation~\cite{SCOR,META} between the magnitude of this splitting and the 
strength of metastabilities detected in the theory at much too coarse lattice 
spacings~\cite{QUE}.}, despite the fact that on general ground they are expected to be 
O($a^2$) corrections~\cite{FR1,FR2}. This is in striking contrast with the smallness of the
cutoff effects observed not only in the mass of the charged pions, which are related 
through the Goldstone theorem to exactly conserved lattice currents~\cite{FMPR}, but 
also in all the other so far measured hadronic observables. Quite remarkably small lattice 
artifacts are found even in matrix elements where the neutral pion is 
involved~\cite{PROC,ETMC1}. 

In this talk, relying on arguments based on the Symanzik analysis~\cite{SYM} of 
lattice cutoff effects, we give an explanation for the origin of such peculiar corrections, 
showing that they are a general feature of any Wilson-like fermion regularization, 
whether twisted or not, and their appearance in the pion mass or instead in other 
observables depends on the choice of the twisted angle (zero or $\pi/2$) and the 
way the critical mass is determined. 

In sect.~\ref{sec:SYMEXP} we recall the properties of the Symanzik approach 
for the description of cutoff effects in LQCD with Wilson fermions and we 
discuss how the critical mass is determined. In sect.~\ref{sec:PMMTM} we 
illustrate the nature of O($a^2$) artifacts in Mtm-LQCD and in sect.~\ref{sec:SWF} 
how they show up in the standard Wilson fermion regularization. We end with 
some concluding remarks in sect.~\ref{sec:CONC}.

\section{Symanzik expansion and critical mass in Wilson fermion LQCD}
\label{sec:SYMEXP}

A) We consider $N_f=2$ LQCD with quarks regularized as Wilson fermions. 
For generic values of the bare (twisted, $\mu$, and untwisted, $m_0$) 
mass parameters the lattice action reads 
\beqn
S_L=S^{\rm YM}_L+\bar\chi\left[\gamma\cdot\widetilde{\nabla}-\frac{a}{2}\nabla^*\nabla
+c_{\rm SW}\frac{ia}{4}\sigma\cdot F+m_0+i\mu\gamma_5\tau^3\right]\chi\, ,
\label{Latact}
\eeqn
where for the sake of generality we have also introduced the clover term.
In this talk we are interested in two specific cases comprised in~(\ref{Latact}).

$\bullet$ Mtm-LQCD, which is obtained from~(\ref{Latact}), by setting $\mu = {\rm O}(a^0)$ 
and $m_0 = M_{\rm cr}^{e}$, where $M_{\rm cr}^e$ is some estimate of the critical mass. 
The physical interpretation of this regularization is most transparent in the 
so-called ``physical basis'', resulting from the field transformation 
\beqn
&&\hspace{-.2cm}\psi=\exp(i\pi\gamma_5\tau^3/4)\chi\, ,\quad 
\bar\psi=\bar\chi\exp(i\pi\gamma_5\tau^3/4)\,\Longrightarrow
\label{CTPB}\\
&&\hspace{-.2cm}S_L^{\rm Mtm} = S^{\rm YM}_L + 
\bar\psi \left[\gamma \!\cdot\! \widetilde{\nabla} -i\gamma_5\tau^3 
\left(- \frac{a}{2} \nabla^*\nabla 
+c_{\rm SW}\frac{ia}{4}\sigma\!\cdot\!F+M_{\rm cr}^{e}\right)+\mu\right]\psi\, .
\label{LatactMtm}
\eeqn 

$\bullet$ The clover standard Wilson fermions action, $S_L^{\rm cl}$, which is obtained 
by setting $\mu=0$ and $m_0 = m+M_{\rm cr}^{e}$ with $m$ an ${\rm O}(a^0)$ quantity. 
With this choice the most appropriate basis for discussing physics is the 
$\chi$-basis itself in which eq.~(\ref{Latact}) was written in the first place.

B) The Symanzik effective Lagrangian associated to the Wilson LQCD 
action~(\ref{Latact}) reads
\beqn 
&&{\cal L}_{\rm Sym} = {\cal L}_4 + \delta {\cal L}_{\rm Sym} \, ,\label{LELgen}\\
&&{\cal L}_4 = {\cal L}^{\rm YM} +\bar\chi[D+m+i\gamma_5\tau^3\mu]\chi 
\, , \qquad  \quad  \delta {\cal L}_{\rm Sym} = a {\cal L}_5 + a^2 {\cal L}_6 + {\rm O}(a^3)  \, ,
\label{L4gen} 
\eeqn 
where the four-dimensional operator (all the necessary logarithmic factors are understood)
specifies the continuum theory in which correlators are evaluated. The very 
definition of effective action, as a tool to describe the $a$ dependence of 
lattice correlators, implies that the mass parameters in ${\cal L}_4$, if not 
exactly vanishing, must be O($a^0$) quantities. Thus all the lattice artifacts 
affecting $M_{\rm cr}^e$ will be described by operators of the form 
$a^{k}\Lambda_{\rm QCD}^{k+1}\bar\chi\chi$, $k=1,2,...$ in $\delta {\cal L}_{\rm Sym}$.

After using the equations of motion of ${\cal L}_4$, the O($a$) piece of 
$\delta {\cal L}_{\rm Sym}$ reads ($b_{5;{\rm SW}}$ and $\delta_1$ are O(1) coefficients)
\beq 
{\cal L}_5 = b_{5;{\rm SW}} \bar\chi i \sigma \cdot F \chi + 
\delta_1 \Lambda_{\rm QCD}^2\bar\chi \chi + {\rm O}(m,\mu) \, . 
\label{L5gen} 
\eeq  
The terms multiplied by powers of $m$ and/or $\mu$ are not specified in 
eq.~(\ref{L5gen}) because they are not of relevance for the topic discussed in this note.  
We recall that the coefficient $b_{5;{\rm SW}}$ vanishes, if $c_{\rm SW}$ in eq.~(\ref{Latact}) 
is set to the value appropriate for Symanzik O($a$) improvement. 

The O($a^2$) part of $\delta {\cal L}_{\rm Sym}$ has a more complicated expression of the type 
\beq 
{\cal L}_6 = \sum_{i=1}^3 b_{6;i} \Phi_{6;i}^{\rm glue} + b_{6;4} \bar\chi \gamma_\mu (D_\mu)^3 \chi + 
\sum_{i=5}^{14} b_{6;i} \Phi_{6;i} + \delta_2 \Lambda_{\rm QCD}^3 \bar\chi \chi + {\rm O}(m,\mu) \, ,
\label{L6gen} 
\eeq 
where the first three operators are purely gluonic, the fourth is a chiral (but not Lorentz) 
invariant fermionic bilinear and the remaining ones are four fermion operators, 
which we find useful to write in the form (equivalence with the list in~\cite{SW} 
can be proved using Fierz rearrangement)
\beq\begin{array}{lll}
& \Phi_{6;5} = (\bar\chi \chi) (\bar\chi \chi) \, , \quad  
& \Phi_{6;6} = \sum_b(\bar\chi \tau^b \chi) (\bar\chi \tau^b \chi) \, ,  
\nonumber \\ 
& \Phi_{6;7} = -(\bar\chi \gamma_5 \chi) (\bar\chi \gamma_5 \chi) \, , \quad  
& \Phi_{6;8} = -\sum_b(\bar\chi \gamma_5 \tau^b \chi) (\bar\chi \gamma_5 \tau^b \chi) \, ,  
\nonumber \\ 
& \Phi_{6;9} = (\bar\chi \gamma_\lambda \chi) (\bar\chi \gamma_\lambda \chi) \, , \quad  
& \Phi_{6;10} = \sum_b(\bar\chi \gamma_\lambda \tau^b \chi) (\bar\chi \gamma_\lambda \tau^b \chi) \, ,  
\nonumber \\ 
& \Phi_{6;11} = (\bar\chi \gamma_\lambda \gamma_5 \chi)  
               (\bar\chi \gamma_\lambda \gamma_5 \chi) \, , \quad  
& \Phi_{6;12} = \sum_b(\bar\chi \gamma_\lambda \gamma_5 \tau^b \chi)  
               (\bar\chi \gamma_\lambda \gamma_5 \tau^b \chi) \, ,  
\nonumber \\ 
& \Phi_{6;13} = (\bar\chi \sigma_{\lambda\nu} \chi)  
               (\bar\chi \sigma_{\lambda\nu} \chi) \, , \quad  
& \Phi_{6;14} = \sum_b(\bar\chi \sigma_{\lambda\nu} \tau^b \chi)  
               (\bar\chi \sigma_{\lambda\nu} \tau^b \chi) \, . \label{4FinL6} 
\end{array} 
\eeq

\indent C) Both in the case of standard Wilson and twisted mass fermions  
the condition to determine the critical mass is the vanishing of 
the PCAC mass. Let us examine these two cases separately. 

1) tm-LQCD -- The condition (it is convenient to rotate the quark fields to the 
$\psi$-basis~(\ref{CTPB}))
\beq
a^3 \sum_{\vec x} \langle 
(\bar\psi \gamma_0 \tau^2 \psi)(\vec x,t)
(\bar\psi \gamma_5 \tau^1 \psi)(0) \rangle\Big{|}_L =0 
\label{LATCOR}
\eeq
leads to a determination of the critical mass which is ``optimal'' 
($M_{\rm cr}^{\rm opt}$) in the sense that with this choice 
all the leading chirally enhanced cutoff effects are eliminated from 
lattice correlators~\cite{FMPR}. 

In the spirit of the Symanzik approach the condition~(\ref{LATCOR}) must be viewed as a 
relation holding true parametrically for generic values of $a$ (and $\mu$). As a result, it 
is equivalent to an infinite set of equations, each equation corresponding 
to the vanishing of the coefficient of the term $a^k$, $k=0,1,2,...$. 
>From the vanishing of the $a^0$ term one gets
\beq 
\int d^3x\, \langle (\bar\psi \gamma_0 \tau^2 \psi)(\vec x,t)
(\bar\psi \gamma_5 \tau^1 \psi)(0) \rangle\Big{|}_{\rm cont} =0\, , 
\label{LATCORC}\eeq
by which restoration of parity and isospin is enforced. This means that, 
if (the O($a^0$) piece of) $m_0$ is chosen so as to verify 
eq.~(\ref{LATCOR}), then we will simultaneously have $m=0$ in~(\ref{L4gen})
and the identification on the lattice of $\bar\psi\gamma_0\tau^2\psi$
with the (time component of the) vector current $V_0^2$ (the identification of 
$\bar\psi\gamma_5\tau^1\psi$ with the pseudoscalar density $P^1$ being trivial). 
The further implications of eq.~(\ref{LATCOR}) are conveniently exposed 
looking at its Symanzik expansion.  
In ref.~\cite{FMPR} it was proved that at O($a$)~(\ref{LATCOR}) implies the condition 
\beqn
&&\xi_\pi \equiv a \, \langle\Omega|{\cal L}_5^{\rm Mtm}|\pi^3(\vec 0)\rangle\Big{|}_{\rm cont} 
\, + \, {\rm O}(a^3) = {\rm O}(a\mu) \, + \, {\rm O}(a^3) \, ,\label{OCM}\\
&&{\cal L}_5^{\rm Mtm} = b_{5;SW} \bar\psi \gamma_5\tau^3 \sigma \cdot F \psi 
+ \delta_1 \Lambda_{\rm QCD}^2\bar\psi i \gamma_5 \tau^3 \psi + {\rm O}(\mu) \, .
\label{L5Mtm}
\eeqn
Eq.~(\ref{OCM}) should be read as a constraint fixing $\delta_1$. 
At O($a^2$) the only relevant term~\cite{ETMC1} is the 
one where $V^2_0 P^1$ is inserted with (the integrated density) ${\cal L}_6^{\rm Mtm}$. 
The latter in the $\psi$-basis has the expression 
\beq
{\cal L}_6^{\rm Mtm}={\cal L}_6^{\rm P-even}+
\delta_2\Lambda_{\rm QCD}^3\bar\psi i\gamma_5\tau^3\psi+{\rm O}(\mu^2) \, ,\label{L6}
\eeq
with ${\cal L}_6^{\rm P-even}$ parity-even. Since in the continuum limit 
(because of parity invariance) one gets $\int d^3x \int d^4y \langle{\cal L}_6^{\rm P-even}(y)
V_0^2(x) P^1(0) \rangle|_{\rm cont}=0$, 
the condition implied by~(\ref{LATCOR}) yields $\delta_2=0$, owing to  
$\int d^3x\int d^4y\langle\bar\psi i \gamma_5\tau^3\psi(y)V_0^2(x) P^1(0)\rangle|_{\rm cont}\neq 0$.
It follows from this analysis is that the estimate of the 
critical mass provided by~(\ref{LATCOR}) is not affected by O($a^2$) effects.
These arguments can be generalized to all orders in $a$ and show 
that $M_{\rm cr}^{\rm opt}$ can only display O($a^{2p+1}$), $p=0,1,...$ 
corrections. The latter are determined by constraints, like~(\ref{OCM}), that 
fix the value of the coefficients $\delta_{2p+1}$ in front of $\bar\psi i \gamma_5\tau^3\psi$.

2) Standard clover Wilson fermions -- The condition for the vanishing of the PCAC mass is 
\beq
\frac{\tilde\partial_0\sum_{\vec x}\langle A_0^b({\vec x,t})P^b(0)\rangle}
{2\sum_{\vec x}\langle P^b({\vec x,t})P^b(0)\rangle}
\Big{|}_L \equiv m_{\rm PCAC}\Big{|}_L=0\quad @ \mu=0\, ,
\label{MPCAC}\eeq
which apart from the normalization and a trivial time derivative is exactly 
eq.~(\ref{LATCOR}) (though written in the $\chi$-basis) with the only difference 
that now $\mu = 0$. This condition is in practice implemented by looking for the 
limiting value of $m_0$ for which $m_{\rm PCAC}\to 0^+$.

The vanishing of the twisted mass is at the origin of all the differences resulting 
from the two ways of subtracting the Wilson term. In fact, if $\mu$ is set to zero, 
from the symmetries of the Wilson theory and the associated Symanzik expansion 
one cannot conclude anymore that $\delta_2$ vanishes. Rather at O($a^2$) 
eq.~(\ref{MPCAC}) fixes the value of $\delta_2$ through the condition 
\beq
\langle\pi(\vec 0)|{\cal L}^{\rm cl}_6|\pi(\vec 0)\rangle\Big{|}_{\rm cont}=0\, ,
\label{SWC}\eeq
where ${\cal L}^{\rm cl}_6$ is the full six-dimensional operator of the Symanzik 
Lagrangian associated to the clover improved Wilson fermion regularization
(including the contribution of the two fermion operator $a\delta_2\Lambda_{\rm QCD}^3\bar\chi\chi$).
In general discretization errors of any order in $a$ will affect 
the critical mass determination~(\ref{MPCAC}) (except those linear 
in $a$ owing to clover improvement).

\section{Neutral and charged pion mass in Mtm-LQCD}
\label{sec:PMMTM}

$\bullet$ {\it Neutral pion mass} -- 
The quantity of interest for the study of 
the neutral pion mass is the zero-momentum four-dimensional Fourier 
transform of the two-point (subtracted) correlator 
\beq
\Gamma_L(p)=a^4\sum_x e^{ipx}\langle P^3(x)P^3(0)\rangle\Big{|}_L\, .
\label{P3P3}
\eeq 
It is immediate to recognize that at $p=0$ and in the limit of very small 
lattice pion mass one gets
\beqn
\Gamma_L(0)=\frac{|G_{\pi^3}|^2}{m^2_{\pi^3}}\Big{|}_L\, ,\qquad
G_{\pi^3}\Big{|}_L=\langle\Omega|P^3(0)|\pi^3(\vec 0)\rangle\Big{|}_L\, .\label{GP3}
\eeqn
>From the Symanzik expansion of $\Gamma_L(0)$ through orders $a^2$ included one can 
prove~\cite{ETMC1} that, even in the absence of the clover term, thanks to the optimal 
choice of the critical mass (see sect.~\ref{sec:SYMEXP}), one arrives at the equation 
\beqn
\frac{|G_{\pi^3}|^2}{m^2_{\pi^3}}\Big{|}_L=\frac{|G_\pi|^2}{m^2_{\pi}}\Big{|}_{\rm cont}\Big{(}1-
a^2\frac{\langle\pi^3(\vec 0)|{\cal L}_6^{\rm Mtm}|\pi^3(\vec 0)\rangle}
{m_{\pi}^2}\Big{|}_{\rm cont}\Big{)}+{\rm O}(\frac{a^2}{m_{\pi}^2})\, ,
\label{NPM}\eeqn
where the continuum pion mass has been simply indicated by $m_\pi$. 
Consistently with the results of $\chi$PT~\cite{LATTCHIR,MUSC,SCOR,SHWU,SSS}, 
a simple Taylor expansion leads to the key formulae of this note   
\beqn
\hspace{-0.4cm}m^2_{\pi^3}|_L=m^2_{\pi}+ a^2\zeta_\pi+{\rm O}(a^2m_{\pi}^2,a^4)\, ,\quad
\zeta_\pi\equiv\langle\pi^3(\vec 0)|{\cal L}_6^{\rm Mtm}|\pi^3(\vec 0)\rangle|_{\rm cont}\, ,\quad
{\cal L}_6^{\rm Mtm}={\cal L}_6^{\rm P-even}\, .
\label{DMASS}\eeqn

$\bullet$ {\it Estimating O($a^2$) lattice artifacts in $m^2_{\pi^3}|_L$} -- 
To estimate the size of the O($a^2$) artifacts~(\ref{DMASS}) we need to compute 
$\zeta_\pi$ in the chiral limit. This can be done 
under the assumption that a sufficiently accurate estimate of $\zeta_\pi$ 
can be obtained in the vacuum saturation approximation (VSA). Quenched 
studies show that VSA works quite well for matrix elements of four-fermion 
operators between pseudo-scalar states~\cite{DON}. We must then identify the 
operators in~(\ref{4FinL6}) that have non-vanishing matrix elements between 
$\pi^3$ states as $m_\pi\to 0$ and give a non-zero contribution in the VSA. 
An example is $P^3P^3= (\bar\psi\gamma_5\tau^3\psi)\,(\bar\psi\gamma_5\tau^3\psi)$ 
which corresponds in the list~(\ref{4FinL6}) to the operator 
$-(\bar\chi\chi)\,(\bar\chi\chi)$. Noticeably one can prove that 
the matrix elements between $\pi^3$ states of the four-fermion operators 
in ${\cal L}_6^{\rm Mtm}$ of interest for our evaluation of $\zeta_\pi$ are all 
proportional to $\langle\pi^3(\vec 0)|{P^3P^3}|\pi^3(\vec 0)\rangle|_{\rm cont}$ 
in the limit $m_\pi^2\to 0$. Thus up to a numerical factor in the VSA we can write 
$a^2\zeta_\pi\sim a^2 |\hat{G}_\pi|^2$, with $\hat{G}_\pi$ the continuum (renormalized) 
analog of the quantity defined in~(\ref{GP3}).

An estimate of $\hat{G}_\pi$ can be obtained either 
by a direct lattice measurement of $a^2{G}_\pi$~\cite{LET} or exploiting the~WTI
$2\hat m_q\langle\Omega|\hat P^3|\pi^3\rangle|_{\rm cont} = f_\pi m_\pi^2$.
Using the results of~\cite{LET,ETMC1}, the two evaluations turn out to be numerically well 
consistent yielding $|\hat{G}_\pi|^{2}\sim (570~{\rm MeV})^4$, a number $\sim 20-25$ times larger than 
the typical scale $\Lambda_{\rm QCD}^4\sim (250~{\rm MeV})^4$. 

$\bullet$ {\it Charged pion mass} -- Replacing the isospin index~3 in eq.~(\ref{P3P3}) 
with either~1 or~2, one finds 
\beqn
m^2_{\pi^\pm}|_L=m^2_{\pi^\pm}+
a^2\langle\pi^\pm(\vec 0)|{\cal L}_6^{\rm Mtm}|\pi^\pm(\vec 0)\rangle|_{\rm cont}
+{\rm O}(a^2{m_{\pi}^2},{a^4})=m_{\pi}^2+{\rm O}(a^2m_{\pi}^2,a^4)\, .\label{MASSC}\eeqn 
The last equality follows from the invariance of ${\cal L}^{\rm Mtm}_{\rm Sym}$ 
under SU(2)$_{\rm ob}\equiv (Q_A^1,Q_A^2,Q_V^3)$ and it is in perfect agreement 
with $\chi$PT~\cite{SCOR,SHWU,SSS} and the similar result derived in ref.~\cite{FMPR}.
The lattice square pion mass splitting in Mtm-LQCD can thus be estimated 
up to terms of O($a^2m_\pi^2,a^4$) with the result   
\beqn
&&\Delta m_\pi^2\Big{|}_{L}^{\rm Mtm}=m_{\pi^3}^2\Big{|}_{L}-m_{\pi^\pm}^2\Big{|}_{L}\sim
a^2\zeta_\pi\sim a^2(570~{\rm MeV})^4 \nn\\
&&\sim (140~{\rm MeV})^2\,\, {@} \,\,a^{-1}\sim 2.3\, {\rm GeV}\, .
\label{MSCN}
\eeqn
This number compares very nicely with the value of the splitting 
$(180(40)~{\rm MeV})^2$ reported in~\cite{LET}.

$\bullet$ {\it Where else does $\zeta_\pi$ enter?} -- Given the impact we have 
seen it has on the lattice expression 
of the neutral pion mass, an important question to ask is where else (besides 
$m_{\pi^3}^2|_L$ and all related energy factors) can the key parameter 
$\zeta_\pi$ appear in the Symanzik expansion of lattice quantities.
The answer requires a detailed analysis which we have no 
space to report here~\cite{ETMC1}. The outcome of it is that to all practical purposes the 
only interesting place where $\zeta_\pi$ enters is just $m_{\pi^3}^2|_L$.

\section{Pion mass and O($a^2$) artifacts with standard Wilson fermions}   
\label{sec:SWF} 

Proceeding as before, one gets for the (clover improved) standard Wilson pions the formula 
\beqn
m_{\pi}^2|_L = m_{\pi}^2 
+ a^2\langle\pi(\vec 0)|{\cal L}_{6}^{\rm cl}|\pi(\vec 0)\rangle|_{\rm cont}
+{\rm O}(a^2m_\pi^2,a^4)\, ,
\label{WPM}\eeqn
where isospin indexes are understood owing to the SU(2) flavour symmetry 
of the lattice Wilson theory. No O($a$) terms are present 
($b_{5;SW}=\delta_1=0$ in~(\ref{L5gen})) as we are assuming clover improvement. 

The particular way in which the critical mass is fixed reflects itself into the form 
of the Symanzik effective Lagrangian of the lattice theory. Here we will 
examine two choices. The first corresponds to the 
standard procedure where the critical mass is determined from eq.~(\ref{MPCAC}). 
The second is somewhat more exotic and corresponds to fixing the critical 
mass by using the determination provided by Mtm-LQCD. 

$\bullet$ {\it The standard way of fixing the critical mass} -- 
At O($a^2$) the condition for the vanishing of $m_{\rm PCAC}|_L$ implies the relation~(\ref{SWC}),  
which fixes $\delta_2$ 
in terms of other parameters of the theory and in particular of the matrix elements of the 
four-fermion operators~(\ref{4FinL6}) between one-pion states. 

Given the uniqueness of the Symanzik expansion (eqs.~(\ref{LELgen}), (\ref{L4gen})) for Wilson 
fermions (close to the chiral limit), a theoretical analysis similar to that we have 
sketched in sect.~\ref{sec:PMMTM}, together with the numerical estimate of $\hat G_\pi$, 
shows that there are sizable contributions in eq.~(\ref{SWC}). As a result chances 
are that $\delta_2\gg 1$ because this coefficient has to compensate for the large 
value of $\zeta_\pi$. The consequences of this situation are twofold. 

1) No O($a^2$) artifacts will affect the value of the lattice pion 
mass because they are absent in $m_{\rm PCAC}|_L$ thanks to~(\ref{SWC}) and,
as it follows by taking the limit $t\to\infty$ in eq.~(\ref{MPCAC}), one has 
\beq
m_{\pi}^2 \frac{f_\pi}{2|G_{\pi}|}\Big{|}_L=m_{\rm PCAC}\Big{|}_L \, . 
\label{MM}
\eeq 

2) On the contrary, in other observables, like for instance the mass of the hadron $h$, 
there will appear O($a^2$) effects proportional to $\langle h|\bar\chi\chi|h\rangle$ 
multiplied by the possibly large coefficient $\delta_2$.

$\bullet$ {\it Using the critical mass of Mtm-LQCD} -- 
If the estimate of the critical mass as determined in Mtm-LQCD is instead employed, 
since, as shown in~\cite{FMPR}, only odd powers of $a$ come into play, the  
term $a^2\Lambda_{\rm QCD}^3\bar\chi\chi$ will not appear in the Symanzik Lagrangian. 
As a result one will have $\delta_2=0$ and the O($a^2$) corrections to the 
square pion mass in eq.~(\ref{WPM}) will not be zero. But now, no O($a^2$) 
corrections stemming from $a^2\Lambda_{\rm QCD}^3\bar\chi\chi$ will affect other 
observables.  

\section{Concluding remarks}   
\label{sec:CONC} 

In this talk we have argued that there are peculiar O($a^2$) cutoff effects in 
LQCD with Wilson fermions which have a dynamical origin related to the mechanism of 
spontaneous chiral symmetry breaking. In Mtm-LQCD they only affect the neutral pion 
mass making it substantially different from that of the charged pion. If the standard Wilson 
fermion regularization is employed, where these discretization errors will show up will depend 
on the way the critical mass is determined. With the usual determination, pion masses 
are free from these lattice artifacts, but the latter will appear 
in other physical quantities, such as hadronic masses. If, instead, the critical mass 
as determined in tm-LQCD is employed such O($a^2$) terms will only affect the value 
of the pion mass.

\vspace{.3cm}
{\bf Acknowledgments -} We would like to thank P. Weisz for useful discussions and comments
and the Organizers of LAT2007 for the lively atmosphere of the meeting. This work 
was partially supported by the EU Contract MRTN-CT-2006-035482 "FLAVIAnet".


\end{document}